\documentclass{article}
\usepackage{spconf,amsmath,graphicx}
\usepackage[dvipsnames]{xcolor}
\usepackage{makecell}
\usepackage{mathtools}
\usepackage{amsmath}
\usepackage{enumitem}
\setlist{nosep, leftmargin=14pt}

\usepackage{mwe} 

\def\thefigdim{.207}
\def\qualifigsep{-4.1mm}
\def\XS{\xspace}
\DeclareMathAlphabet{\mathb}{OML}{cmm}{b}{it}
\def\sbm#1{\ensuremath{\mathb{#1}}}
  \def\xb{{\sbm{x}}\XS}
  \def\yb{{\sbm{y}}\XS}
\def\Fb{{\sbm{F}}\XS}  \def\fb{{\sbm{f}}\XS}
\def\kb{{\sbm{k}}\XS}
\def\db{{\sbm{d}}\XS}
\title{Density Compensated Unrolled Networks for Non-Cartesian MRI Reconstruction}
\name{Zaccharie Ramzi\textsuperscript{(1, 2, 3)}, Jean-Luc Starck\textsuperscript{(2)} and Philippe Ciuciu\textsuperscript{(1, 3)}}
\address{\textsuperscript{(1)}CEA/NeuroSpin, Bât 145, F-91191 Gif-sur Yvette, France. \\ \textsuperscript{(2)}AIM, CEA, CNRS, Université Paris-Saclay, Université Paris Diderot, Sorbonne Paris Cité, F-91191 \\ Gif-sur-Yvette, France. \\ \textsuperscript{(3)}Inria Saclay Ile-de-France, Parietal team, Univ. Paris-Saclay, France.}

\begin{document}
\ninept
\maketitle
\begin{abstract}
Deep neural networks have recently been thoroughly investigated as a powerful tool for MRI reconstruction. 
There is a lack of research, however,  regarding their use for a specific setting of MRI, namely non-Cartesian acquisitions.
In this work, we introduce a novel kind of deep neural networks to tackle this problem, namely \emph{density compensated unrolled neural networks}, which rely on Density Compensation to correct the uneven weighting of the k-space.
We assess their efficiency on the publicly available {\em fastMRI} dataset, and perform a small ablation study.
Our results show that the \emph{density-compensated unrolled neural networks} outperform the different baselines, and that all parts of the design are needed.
We also open source our code, in particular a Non-Uniform Fast Fourier transform for TensorFlow.
\end{abstract}


\section{Introduction}
Magnetic Resonance Imaging (MRI) is a non-invasive medical imaging technique allowing to probe soft tissues in the human body. In MRI, data is collected in the space of the Fourier coefficients $\yb$ of the anatomical image $\xb$, also called k-space. In recent years, the theory of compressed sensing~\cite{Lustig2007} has allowed to dramatically reduce the time spent in acquisition while increasing the time taken for reconstruction, by reducing the amount of measurements collected in the k-space.
Indeed, the idealized (not taking into account the noise, gradient inaccuracies or $B0$-field inhomogeneities) inverse problem we are trying to solve in MRI can be formulated in the single-coil setting as the following:
\begin{equation}
    \yb = \Fb_{\Omega} \xb
\end{equation}
where $F_{\Omega}$ is the under-sampled Fourier operator.
To solve such a problem, a classical method is to define an optimisation problem, with a data consistency term and a regularisation term $R$ (for example the $L1$ norm of the wavelet coefficients of the image). The optimisation problem reads as follows:
\begin{equation}
    \widehat{\xb} = \underset{\xb}{\text{argmin}} \frac12 \|\yb - \Fb_{\Omega}\xb \|_2^2 + \lambda R(\xb)
\end{equation}
Several issues arise with such a method. First to solve the optimisation problem, we need to use an iterative algorithm with potentially computationally-intensive operators like the Fourier Transform involved in the first term or the Wavelet Transform involved in the regularisation term $R(\xb)$. Second, the regularisation term is not necessarily ideal for MR images as it is fixed and not finely tuned to medical images.
While dictionary learning~\cite{Caballero2014} solves partly the second problem, it does so at an increased computation cost.
Deep learning has therefore  been recently introduced~\cite{Zhu2018, Schlemper2018} to tackle both problems, allowing for a fast inference and learning from data.
However, there is a lack of works tackling the issue of data acquired in a non-Cartesian way in the k-space.
Indeed, many different sampling strategies exist for acquiring data in the k-space.
These strategies must meet hardware kinematic constraints on the gradient system~(gradient magnitude, slew rate) and exhibit different advantages (e.g. robustness to motion, maximal sampling efficiency) and limitations (e.g. sensitivity to B0 inhomogeneities, gradient imperfections). 
Different explanations of how and when to use the two non-Cartesian trajectories considered here, radial and spiral, can be found in~\cite{Bernstein2004AdvancedTechniques}.
Importantly, these two trajectories do not acquire the data on a uniform grid in the k-space, but instead off the grid. 
Hence, it is no longer possible to use the classical fast Fourier transform~(FFT) as the forward operator $F_\Omega$. Instead we consider its Non-Uniform extension called NUFFT.

In this study, we tackle the problem of using deep learning for non-Cartesian MRI reconstruction.
Our contributions are threefold:
(i) we design an unrolled MRI reconstruction network, termed Primal-Dual Net~(PDNet) for non-Cartesian data, the first in the literature with a density compensation~(DC) mechanism that adapts to various sampling densities; (ii)
we open source the implementation of this network and that of the NUFFT in the TensorFlow framework; (iii) we benchmark the PDNet architecture endowed with the DC mechanism against U-net and show on the fastMRI data set that the proposed architecture outperforms its competitors and that the DC mechanism indeed helps get better image quality. 

\section{Related Work}
Some recent works~\cite{Hammernik2018LearningData, Schlemper2018, Eo2018} have introduced designs of unrolled (or cross-domain) neural networks to tackle the problem of Cartesian MRI reconstruction.
However, to the best of our knowledge, only one~\cite{Schlemper} has tried to use the same ideas for non-Cartesian data. 
The difference with our work is that they considered only one trajectory, variable-density sampling, and because this trajectory was particular they did not need to introduce DC.
Another difference with the work conducted in~\cite{Schlemper} is that we use emulated single-coil data, and do not synthesize any phase information. 
The data that we handle is also much bigger ($320 \times 320$ target images in our case, and $192 \times 192$ in theirs).

\section{Model}

\subsection{Cross-domain network}
\begin{figure}[h]
\centering
\includegraphics[width=\columnwidth, height=0.7\columnwidth,clip]{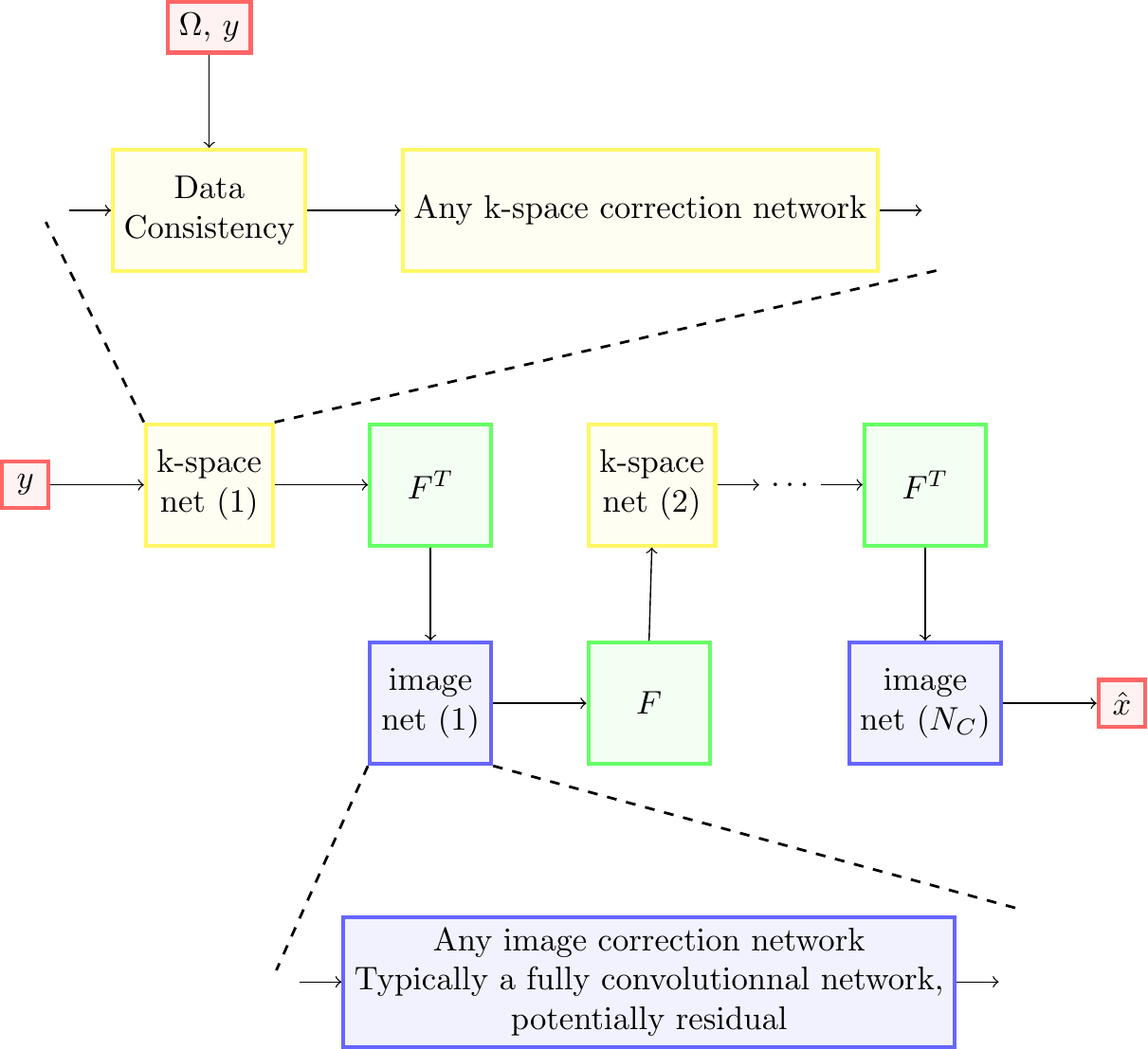}
\caption{
\label{fig:cross-domain} 
General cross-domain networks architecture. Skip and residual connection are omitted for the sake of clarity. $y$ are the under-sampled measurements, in our case the k-space samples, $\Omega$ is the under-sampling scheme, in our case the non-Cartesian trajectories, $F$ is the measurement operator, in our case the Non-Uniform Fast Fourier Transform (NUFFT), and $\hat{x}$ is the recovered solution.
Furthermore, in the particular case of primal only PDNet and with DC, the k-space correction network is simply the application of the previously computed DC factors.
}
\end{figure}

The key intuitive idea behind cross-domain networks is to alternate the correction of the solution of the inverse problem between the measurement space and the image space.
The alternation is done via the use of the measurement operator and its adjoint.
An illustration of cross-domain networks is available in Fig.~\ref{fig:cross-domain}.
The tool to allow that is the unrolling of optimization algorithms~\cite{Gregor2010}.
The Primal-Dual net~\cite{Adler2018}, shortened as PDNet, is derived when unrolling the Primal-Dual Hybrid Gradient optimization algorithm~\cite{Chambolle2010AImaging} (also known as Chambolle-Pock algorithm).
A primal only version also exists, which does not use a k-space correction but simply computes the residual between the original measurements and the measurements associated with the image solution at a given unrolled iteration.
We use this version to avoid using a computationally expensive neural network in the k-space.

\subsection{Non-Uniform Fourier Transform}
The Non-Uniform Fourier Transform is the generalization of the Fourier Transform to positions in the Fourier space that are not necessarily equispaced and on a Cartesian grid.
Let us give the mathematical definition of $\hat{\fb}$, the $d$ dimensional discrete Non-Uniform Fourier Transform of a signal $\fb$, for sample locations $(\xb_i \in [-\frac12; \frac12)^d )_{i=0 \ldots M}$ and frequency indices $(\kb_j)_{j=0 \ldots K}$:
\begin{equation}
     \fb_i = \sum_{j = 0}^K \hat{\fb}_j \exp(-2 i \pi \kb_j^T \xb_i)
\end{equation}
You can find more details about this in~\cite{Keiner2009UsingTransforms}.
An approximate algorithm to have an efficient computation of the Non-Uniform Fourier Transform was introduced in~\cite{Fessler2003NonuniformInterpolation, Beatty2005RapidRatio}.
We refer to this algorithm as the Non-Uniform Fast Fourier Transform (NUFFT), and highlight that unlike the Fast Fourier Transform, it is not an exact algorithm.

\subsection{Data consistency}
The data consistency module of cross-domain networks is the way in which, at every unrolled step, we introduce the information about the original measurements.
Three schemes exist to perform it for the uniform setting. Let us denote the original measurement $\yb$, $\Omega$ the k-space sample locations, $\yb_k$ the full k-space at the $k$-th unrolled iteration, and $M_{\Omega}$ the masking operator. We define the three possible data consistency schemes $f(\yb, \yb_k, \Omega)$ with different (potentially learnable) parameters as follows:
\begin{itemize}
    \item Measurements (soft-)replacement as introduced in Deep Cascade~\cite{Schlemper2018}: At each unrolled iteration, the values of the k-space of the current solution are (soft-)replaced by the original k-space measurements, $$f_{\lambda}(\yb, \yb_k, \Omega) = (1 - M_{\Omega}) \yb_k + (1 - \lambda) M_{\Omega} \yb + \lambda M_{\Omega} \yb_k \, ,$$
    \noindent with $\lambda\in[0,1]$. If $\lambda = 0$, then we have a full replacement. 
    
    \item Learned on the measurements space, as done for the PDNet~\cite{Adler2018}, using convolutionnal networks:
    $$f_{\theta}(\yb, \yb_k, \Omega) = CNN_{\theta}(\yb, M_{\Omega} \yb_k)\, .$$
    \noindent In this case, $f_{\theta}$ is both the data consistency and the k-space correction module.
    \item Measurements residual as done for the primal-only version of the PDNet and in the winning solution of the fastMRI challenge~\cite{Pezzotti2019AnChallenge}: 
    $f(\yb, \yb_k, \Omega) = \yb - M_{\Omega} \yb_k \, .$
\end{itemize}
Out of these three schemes, we chose the measurements residual~(third scheme) to work in the non-Cartesian setting.
Indeed, the replacement scheme does not work because there is no equivalent of the masking operator's supplemental operator (i.e. $1 - M_{\Omega}$) for the non-uniform setting. Also, the learned data consistency would be very computationally intensive. 

\subsection{Density compensation}
Unlike the Cartesian case, the adjoint operator of the NUFFT is not always its inverse operator.
Worse, in most cases, the NUFFT does not admit an inverse operator.
The application of the adjoint operator to the k-space can therefore be very far from the solution to our inverse problem.

To circumvent this, DC has been introduced~\cite{Pipe1999SamplingSolution}.
Indeed, the main problem with the classical MRI trajectories like radial or spiral, is that they over-sample the center of the k-space.
Therefore, when computing the adjoint, a lot of weight is given to a high energy region, resulting in an image with abnormally large values.
DC is just the action of using factors that weigh the different sample locations so that they all play an even role during the application of the adjoint.


In practice, for both the radial and the spiral trajectories, we obtain the DC factors $\db$ by applying the ajoindt and forwar operators iteratively, starting from ones:
\begin{equation}
\label{eq:dc-weights}
\begin{split}
    \db_0 = \mathbf{1}\\
    \db_{n+1} = \frac{\db_n}{F_{\Omega} F^H_{\Omega} \db_n} 
\end{split}
\end{equation}
where the division is here pointwise.

Formally, this means that the data consistency module is rewritten as follows:

\begin{equation}
    f(\yb, \yb_k) = \db_N (\yb -  \yb_k)
\end{equation}

In practice, we take $N=10$.

\section{Experimental Results}

\subsection{Data}
The data used for this benchmark is the emulated single-coil k-space data of the fastMRI dataset~\cite{Zbontar}, along with the corresponding ground truth images. The acquisition was done with 15-channel phased array, in Cartesian 2D Turbin Spin Echo (TSE). The pulse sequences were proton-density weighting, half with fat suppression, half without. The sequence parameters were defined as follows: Echo train length 4, matrix size $320 \times 320$, in-plane resolution 0.5mm$\times$0.5mm, slice thickness 3mm, no gap between slices. In total, there were 973 volumes~(34, 742 slices) for the training subset and 199 volumes~(7135 slices) for the validation subset.

\begin{figure}[h]
\centering
\begin{tabular}{cc}
 
\includegraphics[width=0.35\columnwidth]{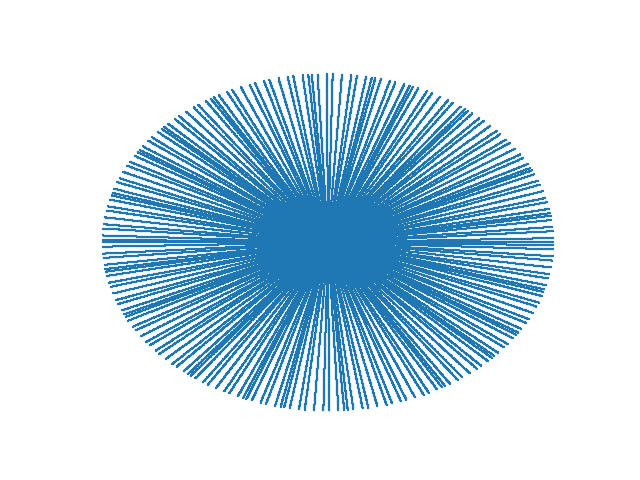}&
\includegraphics[width=0.35\columnwidth]{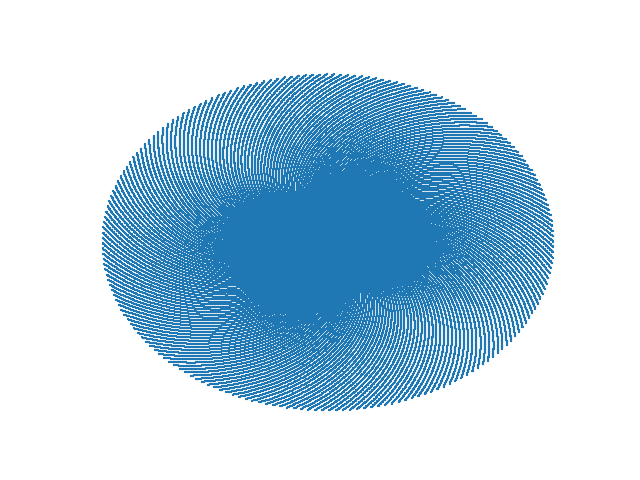} \\
{\bf Radial} & {\bf Spiral}
\end{tabular}
\caption{The 2 k-space trajectories considered in this work. Each of them uses 100 spokes and has a total of $64$k measurements.\label{fig:traj}}
\end{figure}

The full-Cartesian k-space data is then used to reconstruct a complex-valued image.
We then compute the NUFFT of this image to obtain our input data.
The trajectories considered during this work are spiral and radial, with an acceleration factor of 4 compared to the full-Cartesian acquisition and no oversampling was applied.
The trajectories, shown in Fig.~\ref{fig:traj},  are fixed during training and evaluation.

\noindent The study of under-sampled trajectories that are more realistic for TSE, like PROPELLER for radial, is beyond the scope of this paper.

\subsection{Metrics}

The metrics used for evaluating the networks are Peak signal-to-noise ratio~(PSNR) and structural similarity index~(SSIM). They are computed on whole volumes, that is the dynamic range (used for the computation of PSNR and SSIM) is computed on the whole volume. The parameters and definitions used are the same as in the fastMRI paper~\cite{Zbontar}.

\subsection{Experimental setup}

\paragraph*{Implementation.} All the code was done in TensorFlow~\cite{Abadi2016TensorFlow:Systems} and is open sourced\footnote{https://github.com/zaccharieramzi/fastmri-reproducible-benchmark}.
The Non-uniform Fast Fourier Transform in particular is implemented as a TensorFlow package\footnote{https://github.com/zaccharieramzi/tfkbnufft} inspired by the original PyTorch implementation from~\cite{muckley:20:tah}.

\paragraph*{Training details.}
The training loss $L$ was a compound loss inspired by~\cite{Pezzotti2019AnChallenge}:
\begin{equation}
  L(\xb, \hat{\xb}) = \alpha * (1 - MSSIM(\xb, \hat{\xb})) + (1 - \alpha) \|\xb - \hat{\xb}\|_1
\end{equation}
\noindent $\alpha$ was set to 0.98, in order to have the 2 terms of the loss at the same scale. No grid search was carried over this term.
$MSSIM$ is the multiscale structural similarity index defined in~\cite{Wang2003Multi-ScaleAssessment} with the default TensorFlow parameters\footnote{https://www.tensorflow.org/api\_docs/python/tf/image/ssim\_multiscale}.
We used the RADAM optimizer~\cite{Liu2020OnBeyond} with a learning rate of $10^{-4}$ and the default TensorFlow parameters\footnote{https://www.tensorflow.org/addons/api\_docs/python/tfa/optimizers}.
To allow for an easier training, we scale the data by $10^6$ in order to avoid low values that result in unstable training. No grid search was carried over this term.
Each batch is composed of exactly one slice of a whole volume taken at random\footnote{Contrarily to our previous work we do not select only central slices.}. An epoch is defined as seeing exactly one slice of each volume in the training data set, that is 973 slices.

The training was carried for 100 epochs for all the networks on one V100 GPU with 32GB of RAM. For the U-net it lasted 8 hours, while it took 1 day for the PDNet.

\paragraph*{Comparison.} We compared our approach to the following methods: (i) application of the adjoint operator with DC; (ii) primal only PDNet without DC used for k-space correction, but a simple normalization by the maximum value (without any normalization the network does not train at all); (iii) a U-net~\cite{Ronneberger} getting as input the output of the adjoint operator with DC (the U-net was trained residually and used 16 base filters). 

This comparison serves as an ablation study to verify that the main design aspects of the network, unrolling and DC, are critical for image quality.
All the networks were trained in the same way.

\subsection{Results}
\paragraph*{Quantitative results.}

\begin{table}[]
\small
\centering
\caption{Mean PSNR / SSIM on the validation volumes of the different approaches for both contrasts. The best results are in bold font.}
\label{tab:psnr-ssim}
\resizebox{\columnwidth}{!}{%
\small
\begin{tabular}{|l|c|c|c|}
\hline
\textbf{Model}                 & \textbf{Radial} & \textbf{Spiral} & \textbf{\# Parameters}\\ \hline
\textbf{PDNet no DC}           & 27.02 / 0.6747         & 28.02 / 0.6946     & 156k      \\ \hline
\textbf{Adjoint + DC}          & 27.11 / 0.6471           & 31.70  / 0.7213   & 0      \\ \hline
\textbf{U-net on Adjoint + DC} & 32.26 / 0.7224          & 32.82 / 0.7460   &   481k    \\ \hline
\textbf{PDNet w DC}            & \textbf{32.66 / 0.7327}  & \textbf{33.08 / 0.7534}  & 156k \\ \hline
\end{tabular}
}
\end{table}


The quantitative results in Table~\ref{tab:psnr-ssim} show that our approach~(PDNet with DC) outperforms the others.
Moreover, these results show that to obtain good results for these trajectories, combining unrolling and DC is instrumental, as using only one of these two ingredients will lead to degraded performance.

The quantitative results are slightly higher for spiral trajectory because it has a better coverage of the high frequencies in the k-space.


\paragraph*{Qualitative results.}
\begin{figure*}[h]
\begin{tabular}{@{\hspace*{\qualifigsep}}c@{\hspace*{\qualifigsep}}c@{\hspace*{\qualifigsep}}c@{\hspace*{\qualifigsep}}c@{\hspace*{\qualifigsep}}c@{\hspace*{\qualifigsep}}}
\includegraphics[width=\thefigdim\linewidth]{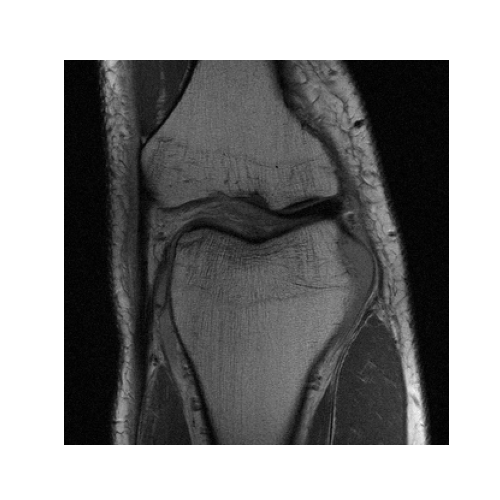}&
\includegraphics[width=\thefigdim\linewidth]{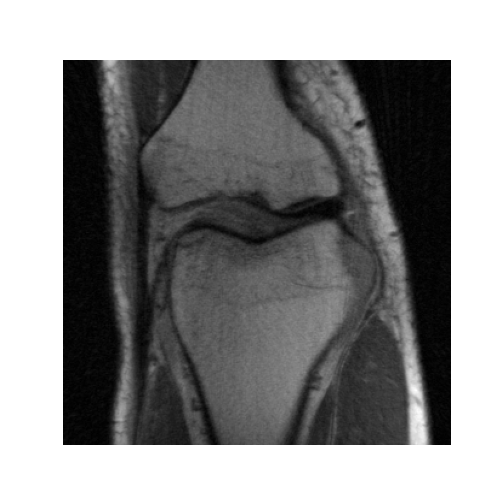}&
\includegraphics[width=\thefigdim\linewidth]{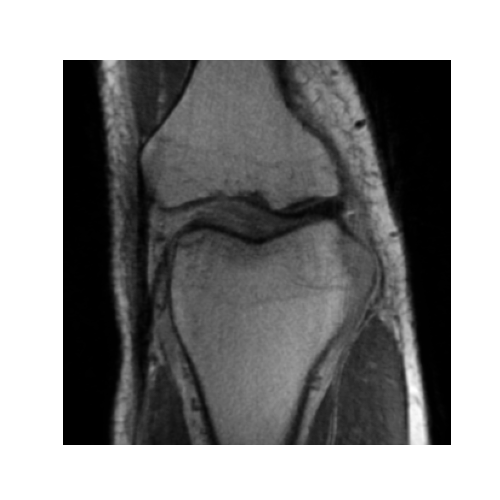}&
\includegraphics[width=\thefigdim\linewidth]{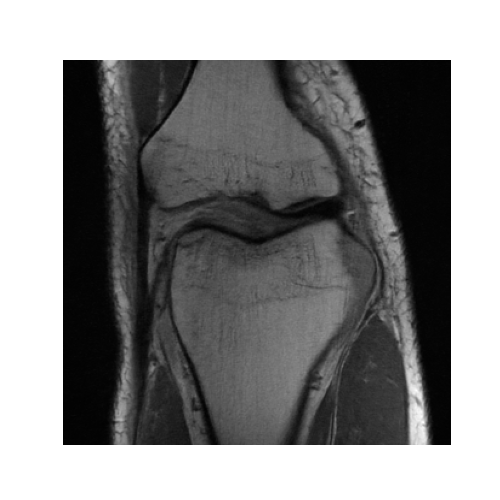}& 
\includegraphics[width=\thefigdim\linewidth]{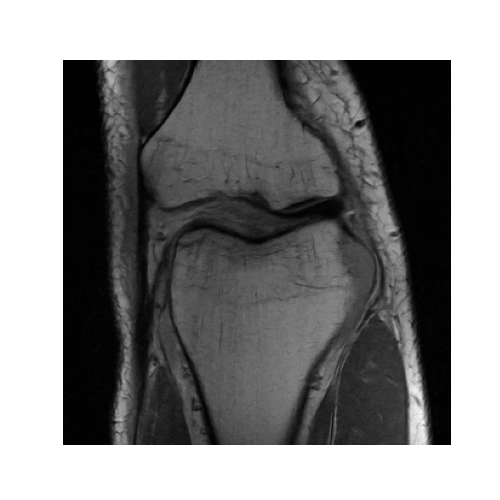}\\[-.85cm]
&
\includegraphics[width=\thefigdim\linewidth]{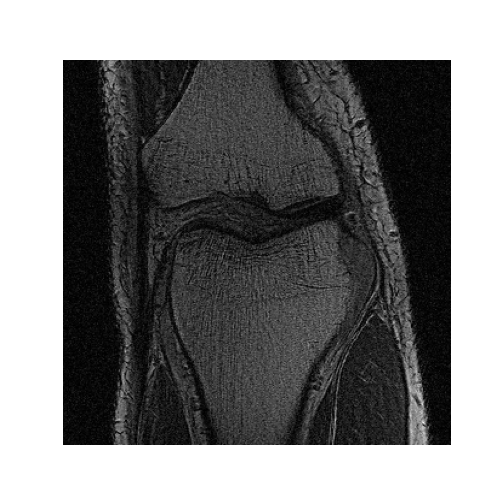}&
\includegraphics[width=\thefigdim\linewidth]{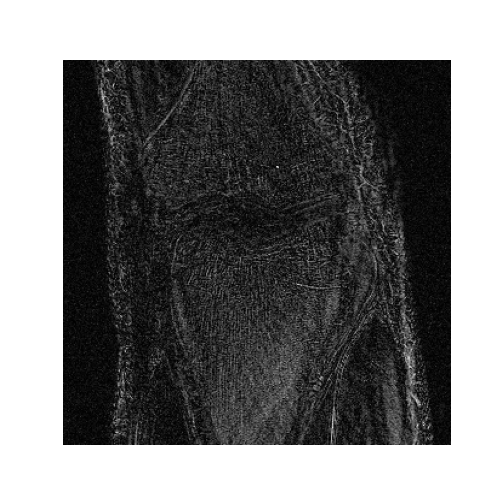}&
\includegraphics[width=\thefigdim\linewidth]{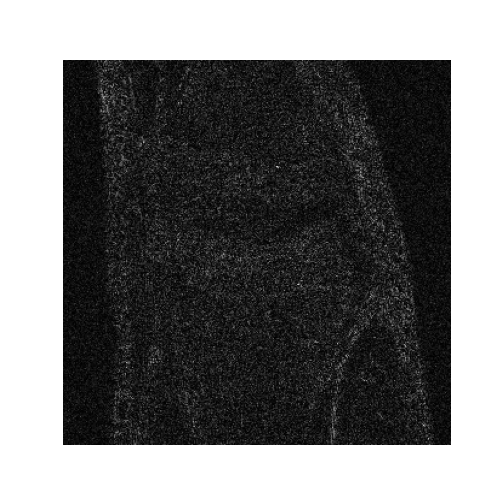}& 
\includegraphics[width=\thefigdim\linewidth]{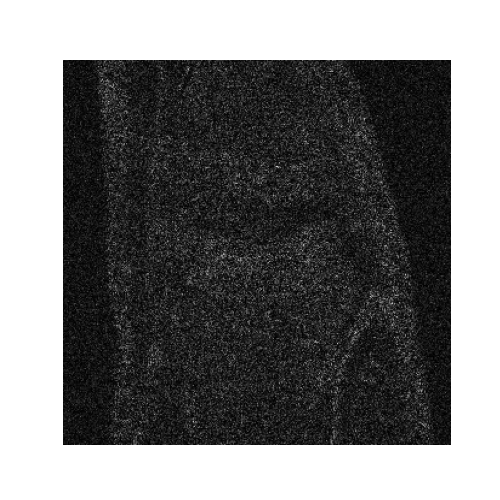}\\[-6.75cm]
{\bf Reference} & {\bf Adjoint + DC} & {\bf PDNet no DC} & {\bf U-net on Adjoint + DC} & {\bf PDNet w DC}\\[6cm]
\end{tabular}
\caption{\textbf{Radial acquisition}: Reconstruction results for a specific slice (16th slice of \texttt{file1001184}, part of the validation set). The first row represents the reconstruction using the different methods, while the second represents the absolute error when compared to the reference. \label{fig:result_reconstruction_radial}}
\end{figure*}

\begin{figure*}[!h]
\begin{tabular}{@{\hspace*{\qualifigsep}}c@{\hspace*{\qualifigsep}}c@{\hspace*{\qualifigsep}}c@{\hspace*{\qualifigsep}}c@{\hspace*{\qualifigsep}}c@{\hspace*{\qualifigsep}}}
\includegraphics[width=\thefigdim\linewidth]{Figures/image_gt.png}&
\includegraphics[width=\thefigdim\linewidth]{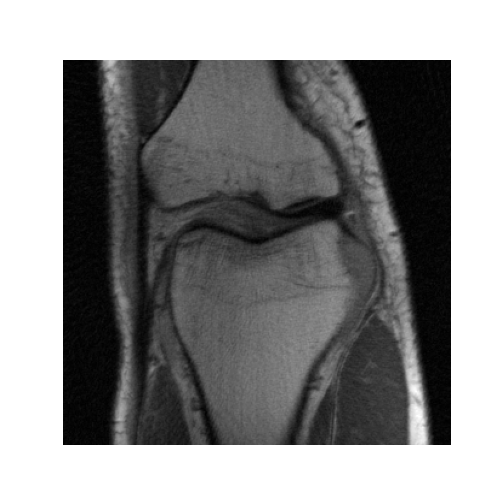}&
\includegraphics[width=\thefigdim\linewidth]{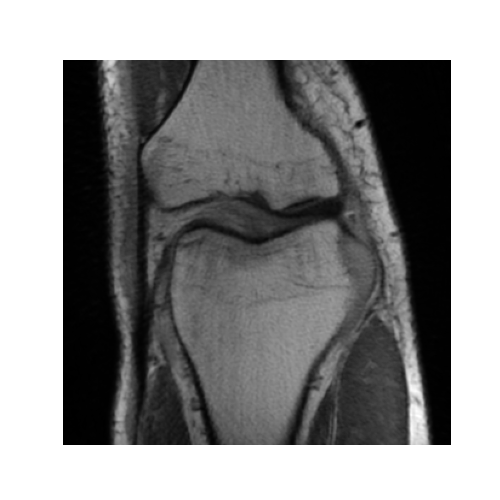}&
\includegraphics[width=\thefigdim\linewidth]{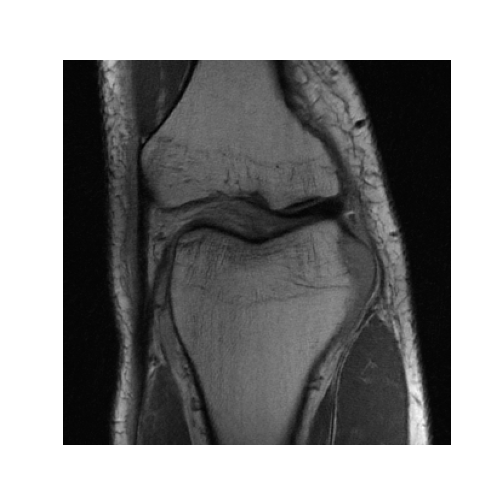}& 
\includegraphics[width=\thefigdim\linewidth]{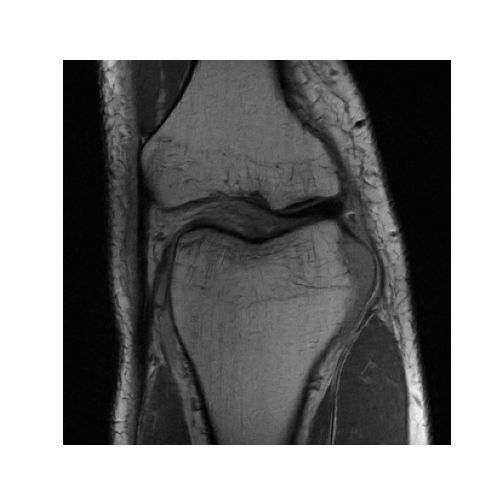}\\[-.85cm]
&
\includegraphics[width=\thefigdim\linewidth]{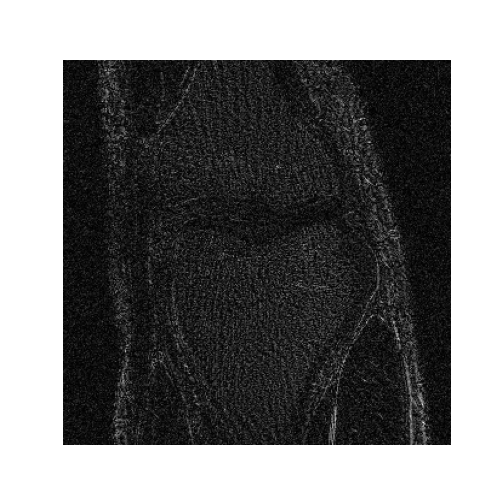}&
\includegraphics[width=\thefigdim\linewidth]{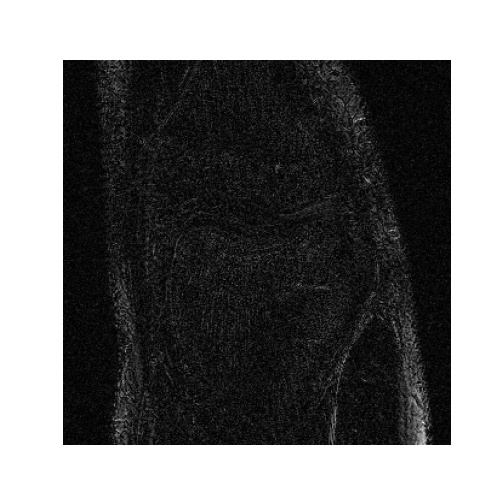}&
\includegraphics[width=\thefigdim\linewidth]{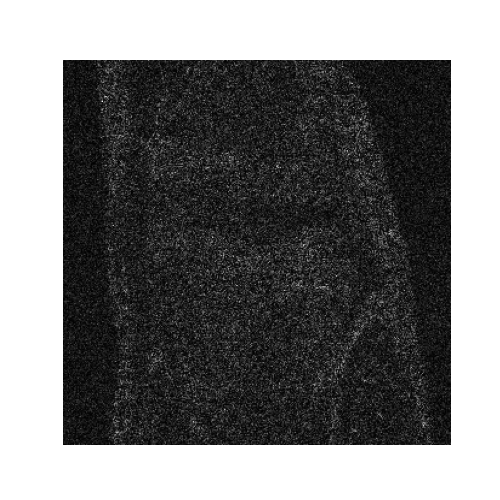}& 
\includegraphics[width=\thefigdim\linewidth]{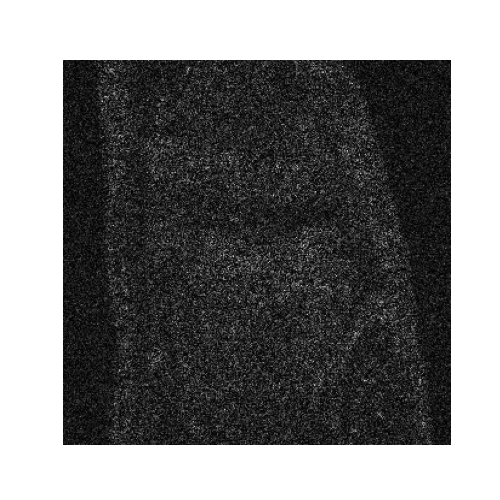}\\[-6.75cm]
{\bf Reference} & {\bf Adjoint + DC} & {\bf PDNet no DC} & {\bf U-net on Adjoint + DC} & {\bf PDNet w DC}\\[6cm]
\end{tabular}
\caption{\textbf{Spiral acquisition}: Reconstruction results for a specific slice (16th slice of \texttt{file1001184}, part of the validation set). The first row represents the reconstruction using the different methods, while the second represents the absolute error when compared to the reference. \label{fig:result_reconstruction_spiral}}
\end{figure*}

The visual inspection of the reconstructed MR images confirm the quantitative measurements.
In particular, one can visualize in Fig.~\ref{fig:result_reconstruction_radial} and~\ref{fig:result_reconstruction_spiral} that the image's inner contrast is better recovered by the PDNet, and the structures are sharper.
Finally, the residual map is slightly flatter for the PDNet with DC compared to the U-net, especially around structures.

\section{Conclusion \& Discussion}
In this work, we described the extension of unrolled networks with the use of density compensation, namely the density-compensated unrolled networks.
Via an ablation study, we show some evidence to support the fact that the elements of this design are needed to reach an acceptable image quality for the considered trajectories.
However, as opposed to Cartesian data, for which the fastMRI challenge is available, there is no public benchmark for non-Cartesian MR image reconstruction.
It is therefore difficult to compare results with other works as the sampling trajectories might differ.

The major future direction of this work is its extension to more complex and realistic reconstruction settings, namely multi-channel phased array coil, 3D imaging and their combination.
This will certainly mean for the models to be implemented in a multi-GPU fashion, placing each image-correction/k-space correction block sequentially on a different GPU.
At some point, the U-net might still be the more viable alternative as there is no need to use the NUFFT on GPU for back-propagation.

The other future direction of this work is to use density compensated unrolled networks for trajectories derived from radial and spiral, like the SPARKLING trajectory~\cite{Lazarus2018}.

\section{Compliance with Ethical Standards}
\label{sec:ethics}

This research study was conducted retrospectively using human subject data made available in open access by~\cite{Zbontar}. 
Ethical approval was not required as confirmed by the license attached with the open access data.

\section{Acknowledgements}
We are grateful to Jo Schlemper for his very useful remarks and answers to our questions. We also thank Chaithya G.R. for the discussions about DC. We acknowledge the financial support of the Cross-Disciplinary Program on Numerical Simulation of CEA for the project entitled SILICOSMIC.
We also acknowledge the French Institute of development and ressources in scientific computing~(IDRIS) for their AI program allowing us to use the Jean Zay supercomputer's GPU partitions.
\bibliographystyle{IEEEbib}
\bibliography{code,main}

\end{document}